\documentclass[twocolumn,showpacs,preprintnumbers,amsmath,amssymb]{revtex4}
\pdfoutput=1

\usepackage{graphicx}
\usepackage{dcolumn}
\newcommand{\be}{\begin{equation}}
\newcommand{\ee}{\end{equation}}
\newcommand{\beq}{\begin{equation}}
\newcommand{\beql}[1]{\begin{equation}\label{#1}}
\newcommand{\eeq}{\end{equation}}
\newcommand{\ba}{\begin{array}}
\newcommand{\ea}{\end{array}}
\newcommand{\bea}{\begin{eqnarray}}
\newcommand{\beal}[1]{\begin{eqnarray}\label{#1}}
\newcommand{\eea}{\end{eqnarray}}
\newcommand{\ben}{\begin{enumerate}}
\newcommand{\een}{\end{enumerate}}
\newcommand{\bean}{\begin{eqnarray*}}
\newcommand{\eean}{\end{eqnarray*}}

\newcommand{\btab}[1]{\begin{tabular}{#1}}
\newcommand{\etab}{\end{tabular}}

\newcommand{\comment}[1]{}

\begin{document}

\title{Statistical thermodynamics and weighted topology of radial networks}

\author{Rak-Kyeong Seong}
\email{rak-kyeong.seong@imperial.ac.uk}
\author{Dimitri D. Vvedensky}
\email{d.vvedensky@imperial.ac.uk}
\affiliation{The Blackett Laboratory, Imperial College London, London SW7 2AZ, United Kingdom}

\date{\today}

\begin{abstract}
We propose a method of characterizing radial networks based on a partition function associated with
the structural triangulation of the network.  The internal energy, Helmholtz free energy, and
entropy derived from the partition function are used to group similar networks together and to
interrogate the history of their development. We illustrate our methodology for a model for optimal
transport in tree leaves.
\end{abstract}


\maketitle

Interest in networks has expanded rapidly in recent years \cite{albert02,boccaletti06,dorogovtsev08}.
In particular, there has been substantial interest in problems associated with modelling venation
patterns \cite{corson10,clarke06,runions05}, optimizing supply networks \cite{katifori10,dodds10},
and determining final system shapes and sizes based on a simulated growth mechanisms \cite{xia07}.

The characteristics of a network are determined by the supply and demand mechanisms of the system.
For example, a leaf is supplied with nutrients and sunlight as energy sources, and demand is created
by the biological drive to keep the leaf cells alive and the leaf growing.  For a river ending in a
delta, the supply is the amount of rain falling, and demand is created by the pull of gravity, the
amount of water dissipation, and the terrain of the delta.

The rules and parameters governing the supply and demand of a system are expected to be invariant
over time scales beyond its lifespan. Hence, the same rules and parameters govern the creation and
the extinction states of a system. What happens between these is affected by external factors. For
example, a leaf may suffer under less than ideal weather conditions or a river may be affected by
the construction of a dam.

An important question is whether systems can be grouped according to their shape and venation at a
given state in their lifespan. Going a step further, one may ask whether interpretations of life
history can be made by analyzing and comparing a state in the lifespan of a system to a comparable
state of another system.  We expect that the growth history of a system is encoded in the
distribution of energy in its venation, i.e.~the work done in building up the supply system. Due to
external factors and the existing structure of the system, some parts of a venation system may have
required more energy to build up than other parts.

In this paper, we introduce a method for measuring the energy distribution of the venation of a
network based on a partition function calculated from its energy states, as represented by a
structural triangulation.  The partition function is used to calculate thermodynamic variables, which
enable systems to be grouped together at their maturation state -- the state when the system stops
growing, as well as enabling the development of a network to be investigated. We illustrate our
methodology by analyzing simulated tree leaves.

Topological invariants are used to characterize shapes and structures of geometrical objects. A
prominent example is the Euler characteristic $\chi$ \cite{nakahara03}, a measure of the
\textit{topological} structure of a space that is invariant under deformations. For convex
polytopes, for example, $\chi=V-E+F=2$ where $V$, $E$ and $F$ are the numbers of vertices, edges
and faces, respectively, of the polytope.

We will focus on planar radial networks with a single origin and without loops. Let the venation of
the system consist of edges $e=(v_i,v_j)$ connecting vertices $v_i=(x_i,y_i)$, with one vertex
designated as the origin $v_0=(0,0)$, which is associated with the first generation of the network,
$g(v_0)=1$. Then, an edge $e=(v_i,v_j)$ always connects an $n$th generation vertex, $g(v_i)=n$, to an
$(n+1)$th generation vertex, $g(v_j)=n+1$.

We define a triangular element $\Delta$ as a set of three vertices, $\Delta=\{v_i,v_j,v_k\}$, such
that the generational association is $g(v_i)=n$ and $g(v_j)=g(v_k)=n+1$. A triangular element is
assigned the generation $g(\Delta)$, which corresponds to the lowest generational number of its
vertices.  We emphasize that there is no intersection between triangular elements.  The set of all
triangular elements in a network is called the triangulation ${\cal M}$ of that network.  Although
we have not proven so, the uniqueness of the triangulation of a network according to the foregoing
prescription is a plausible working hypothesis.

\begin{figure}
\includegraphics[totalheight=9cm]{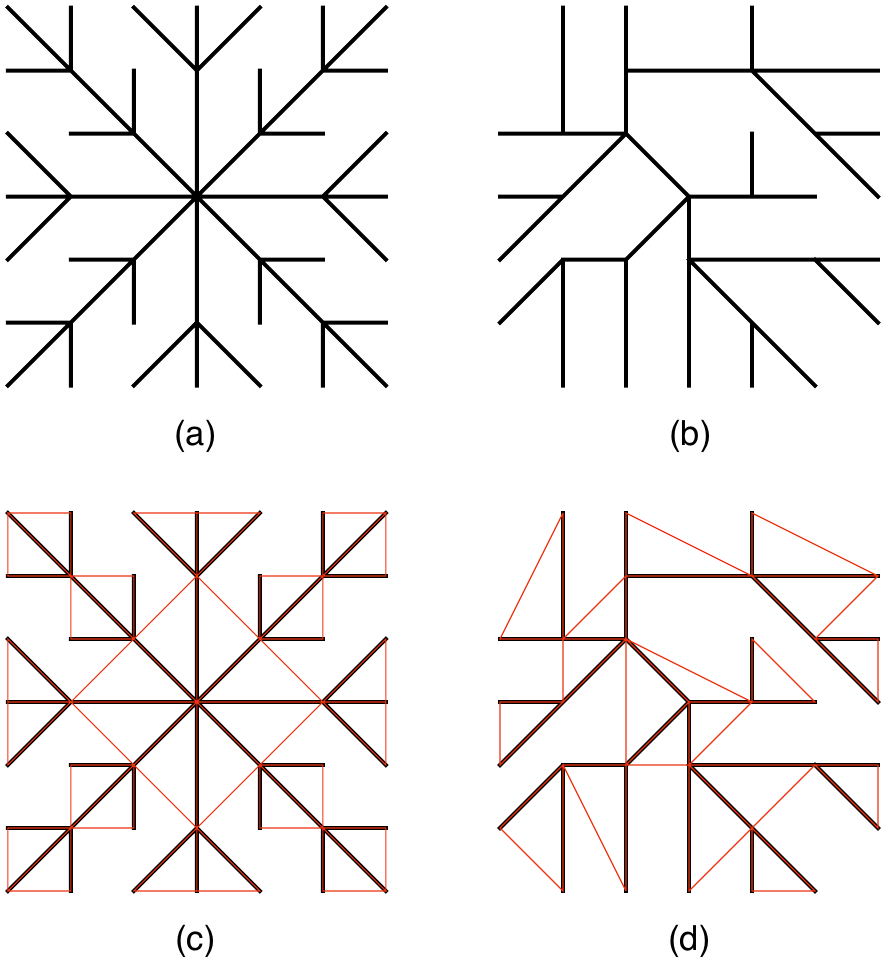}
\caption{(Color online) Two venation networks (upper panels) and their triangulations (lower
panels).}
\label{fig1}
\end{figure}

Two venation networks and their triangulations are shown in Fig.~\ref{fig1}. Each triangular element
$\Delta_i$ not only has an associated generation $g(\Delta_i)$, but also an associated area
$A(\Delta_i)$.  While the structure of a triangulation is unchanged by deformations of the network,
the areas of the elements allow the original and transformed networks to be distinguished. 
Accordingly, we use the generation and area associated with the triangular elements as a measure of
the energy distribution in a venation. In effect, we have a weighted topological description of the 
network based on its triangulation.

For increasing generation number $g$, the system expends correspondingly more energy creating a
vertex $v$ and element $\Delta$ associated with $g$. The system must move material from the supply
source at the origin to the position of $g$. A reasonable physical interpretation for $g$ is,
therefore, an energy level of the system. If the energy unit of the system is $E_0$, the energy
$E_g$ is the work required to produce a unit at generation $g$:~$E_g=g E_0$. Building on this point
of view, a natural interpretation of the area $A(\Delta)$ of an element $\Delta$ is a measure of the
material  needed to assemble this element at generation $g(\Delta)$. In general, if the material
needed for an element of the system is $N$, then $N(\Delta)=\rho A(\Delta)$, where $\rho$ is a
proportionality constant.  Without loss of generality, we take $N$ to be a positive integer, and set
$\rho=1$. This requires the coordinates of the vertices to be integer-valued as well,
$x_i\in\mathbb{Z}$, such that the network is embedded in an integer lattice $\mathbb{Z}^{2}$. This
modification is not a requirement, but it allows us to interpret $N(\Delta)$ as the number of
`particles' with energy $E_{g(\Delta)}$ required to create the triangle element $\Delta$.

We are now able to define the partition function of the triangulation $\mathcal{M}$ as
\begin{equation}
Z(\mathcal{M})=\sum_{\Delta\in\mathcal{M}}{N(\Delta)~t^{g(\Delta)}}\, ,
\label{eq3}
\end{equation}
where $t=e^{-\beta E_{0}}$ and $\beta$ the inverse temperature.  The factor
$t^{g(\Delta)}$ originates from the fact that particles in the same generation are assigned to the
same energy level $E_{g(\Delta)}=g(\Delta)E_0$.  The partition functions for the  networks in
Fig.~\ref{fig1} are
\begin{align}
Z(\mathcal{M}_c)&=16 t + 16 t^2 + 8 t^3\, ,\\
Z(\mathcal{M}_d)&=5 t + 9 t^2 + 9 t^3 + 4 t^4 + t^5\, ,
\label{eq4}
\end{align}
where $\mathcal{M}_c$ and $\mathcal{M}_d$ are the triangulations in Fig.~\ref{fig1}(c) and
\ref{fig1}(d), respectively.

The partition function is the sum over the microstates the system can occupy.  These
are identified in the network as `particles' that have an associated history. To
create a particle in element $\Delta$, the network had to occupy a specific state where work
$E_{g(\Delta)}$ was done to create this particle. Accordingly, when we sum over all network
system particles in (\ref{eq3}), we are indeed summing over all microstates in its history with
associated energy $E_{g(\Delta)}$.  Because energy levels depend on the generation $g$ of the
triangle elements which, in turn, depend on the structure of the venation, the partition function in
(\ref{eq3}) can be considered not only as a statistical measure of the network, but also as a
measure of the structure of the venation. Therefore, we expect two systems with similar partition
functions to display similar venation structures.

\begin{figure}[b!]
\includegraphics[width=9cm]{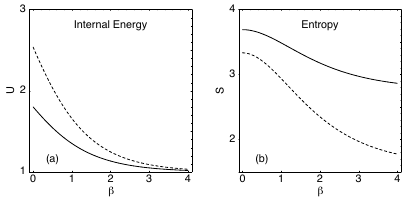}
\caption{(a) Internal energy and (b) entropy for the networks in Fig.~\ref{fig1}(a) (solid
curves) and Fig.~\ref{fig1}(b) (broken curves).}
\label{fig2}
\end{figure}

\begin{figure*}[ht!]
\includegraphics[width=0.9\textwidth]{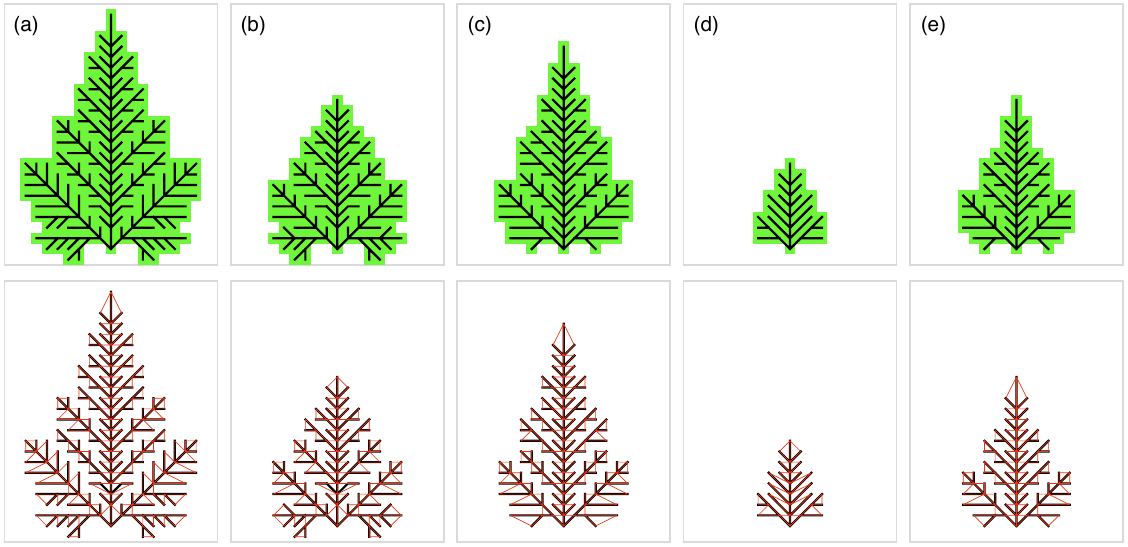}
\caption{(Color online) Simulated leaves (top panels)  with parameters (a) $\alpha=0.6,\beta=0.5$,
(b) $\alpha=0.65,\beta=0.38$, (c) $\alpha=0.6,\beta=0.8$, (d) $\alpha=0.75,\beta=0.7$,  and (e)
$\alpha=0.65,\beta=0.7$, with corresponding triangulations (bottom panels). All leaves have
$\epsilon=4.0$.}
\label{fig3}
\end{figure*}

The partition function enables us to calculate thermodynamic variables for the network.   The
internal energy is
\begin{equation}
U=-\frac{\partial \ln{Z}}{\partial\beta}\, .
\end{equation}
Figure \ref{fig2}(a) shows that $U(\mathcal{M}_b)>U(\mathcal{M}_a)$ for all $\beta$. In the first
few generations (large $\beta$), the internal energies of the two networks are similar, but diverge
as higher generations are included (decreasing $\beta$). The physical interpretation of this is that
there has been more work done by the network system to create the venation of the asymmetric network
in Fig.~\ref{fig1}(b) than that of the symmetric network in Fig.~\ref{fig1}(a). Thus, the structure
of the asymmetric network displays a higher internal energy through more elements being in higher
generations of the system than in the symmetric network. These qualitative trends can already be
seen in the partition functions. In general, we expect that a network which displays a higher degree
of venation symmetry than another network system to have the lower internal energy.

The entropy $S=k_{B}(\ln{Z}+\beta U)$ in the canonical ensemble, where we have set Boltzmann's
constant $k_{B}=1$.  As Fig.~\ref{fig2}(b) shows, $S(\mathcal{M}_a)>S(\mathcal{M}_b)$ for all
$\beta$. To understand this result, we identify entropy with the energy dispersal in the network,
rather than as a measure of structural `disorder' of the venation.  The energy in the symmetric
network is more evenly distributed between the different generations than in the asymmetric
network and, hence, has the higher entropy.  Two limiting forms of the entropy are especially
useful.  As $\beta\to\infty$ ($t\to0$), the entropy approaches the logarithm of the triangulated area
associated with the first generation:
\begin{equation}
\lim_{\beta\to\infty}S=\ln\biggl[N(\Delta)\Big|_{g=1}\biggr]\, .
\label{eq5}
\end{equation}
The other limit is $\beta\to0$ ($t\to1$), where the entropy approaches the logarithm of the area of
the entire triangulation:
\begin{equation}
\lim_{\beta\to0}S=\ln\biggl[\sum_{\Delta\in{\cal M}}N(\Delta)\biggr]\, .
\label{eq6}
\end{equation}
Both of these limits can be obtained directly from the partition function;~the interesting situation
is when the entropy curves of different networks cross.

We consider now a model for leaf growth based on the notion of optimized transport. Supplying a leaf
with water and nutrients has a cost. Minimizing that cost determines the shape of the leaf. We
follow Xia \cite{xia07} in developing a model for the supply network of tree leaves that is
parametrized by three quantities that characterize the cost function associated with extending the
network.  The transport cost is minimized when nutrients are moved in the same direction and
maximized when moved in opposing directions.  We eliminate the option of opposite flows in the
transport network by setting the cost of sustaining such a flow to infinity. For all other flows, the
parameter $\beta$ scales the extent to which the flow direction contributes to the total cost of
sustaining the supply network. Another cost is associated with supply flowing through a network edge.
The more matter that flows through an edge, the larger the cross-sectional area of that edge which
requires more energy to sustain the pressure in the descendant supply network. The `thickness' of a
given edge is proportional to its weight, so the cost of maintaining an edge with a particular weight
leads to a second parameter, $\alpha$.  A third parameter, $\varepsilon$, is an efficiency
coefficient for  photosynthesis.

\begin{figure}[t!]
\includegraphics[width=9cm]{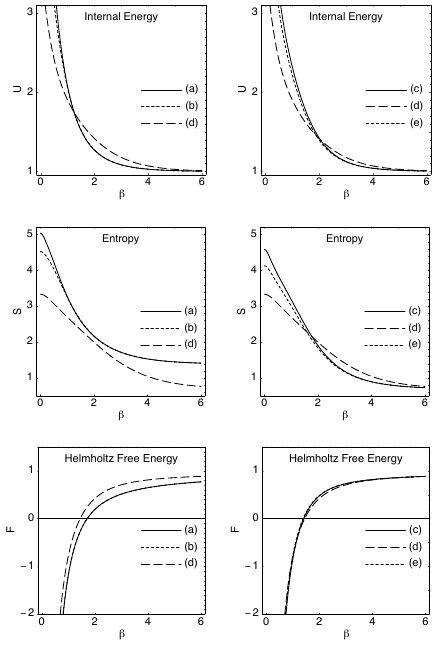}
\caption{The internal energy, Helmholtz free energy, and the entropy for each of the simulated
leaf networks in Fig.~\ref{fig3}.}
\label{fig4}
\end{figure}

Using this model of leaf growth, we have simulated the growth of five tree
leaves, which are shown in Fig.~\ref{fig3}, together with their parameters in the cost function.  The
corresponding thermodynamic functions are shown in Fig.~\ref{fig4}.  The thermodynamic profiles of
the tree leaves have two regimes that are of particular interest. The high energy regime 
($\beta\rightarrow 0$) can be associated with the late stage of the leaf development process, while
the low energy regime ($\beta\rightarrow \infty$)with the early stage of
development.

The internal energy profiles show that, in the early stages of growth ($\beta\rightarrow\infty$),
similar work has been done to `build' all five leaves. In the later stages of development, the
amount of work for leaf (d) becomes the lowest, which agrees with intuitive expectations, since this
leaf is the smallest in the sample.  The entropy and internal energies of leaves (a) and (b) are
essentially indistinguishable during the early stages of growth, as are those of leaves (c) and
(e).  However, discernible differences in these quantities appear at later stages.  Leaf (d) is
different from all other leaves because of its size, and this is also apparent from its
thermodynamic functions.  The entropy of leaf (d), in particular, shows a qualitatively different
profile from all other leaves, which is consistent with our earlier interpretation of entropy as the
distribution of energy between the different generations in a network.

The free energy profiles allow us to classify the sample leaves into two main
groups:~$\{a,b\}$ and $\{c,d,e\}$. This grouping is confirmed by the low-energy limit
($\beta\rightarrow\infty$) of the entropy profiles of the tree leaves, where the entropies of the
tree leaves correspond to the areas of their triangulations in the first generation, as noted in
(\ref{eq5}).  For leaves with the same initial entropy, differences arise when later generations of
the triangulation are included.  This is due to differences in the optimization parameters used for
the simulations.

We have described a method for analyzing networks based on a triangulation that associates an energy
with each generation of the network.  This enables a partition function to be constructed 
from which standard thermodynamic functions can be directly computed.  The interpretation of
these functions for networks provides a powerful classification scheme. The comparison of the
internal energy, entropy, and Helmholtz free energy for several simulated leaves, all of which grow
by the same mechanism, but with differently parametrized cost functions, illustrates how our
methodology is able to discriminate between  networks that show subtle structural differences.  Our
method differs from other thermodynamic formulations of networks \cite{bianconi08,bianconi09} in
that our method characterizes individual networks through their triangulation, rather than requiring
an ensemble of networks.

Perhaps the most far-reaching aspect of our analysis is that we are able to study the history of
a family of networks, especially where the early stages of development are similar, but then deviate
at some later stage, either from an intrinsic mechanism, as shown here, or from an external
influence.  Hence, our method is eminently suitable for examining biological networks, where the
entire developmental history, rather than simply the fully-developed network, has fundamental
importance.

R.-K. S. is grateful for the support of Placental Analytics LLC.

\end{document}